\documentclass[aps,pra,amsmath,twocolumn,showpacs]{revtex4}
\usepackage{amsfonts,amssymb,graphicx,color}
\usepackage{rotating}

\newcommand{\ve}[1]{\langle #1 \rangle}
\newcommand{\Var}{\textrm{Var}}
\newcommand{\var}{\textrm{Var}}
\newcommand{\dg}{^\dagger}

\newcommand{\be}{\begin{equation}}
\newcommand{\ee}{\end{equation}}

\newcommand{\ex}[1]{e^{#1}}
\newcommand{\id}{\openone}

\newcommand{\tr}{{\rm tr}}
\newcommand{\ket}[1]{\left|{#1}\right\rangle}
\newcommand{\bra}[1]{\left\langle{#1}\right|}
\newcommand{\braket}[2]{\langle{#1}|{#2}\rangle}
\newcommand{\ketbrad}[1]{\left|{#1}\rangle\!\langle{#1}\right|}
\newcommand{\ketbra}[2]{\left|{#1}\rangle\!\langle{#2}\right|}
\newcommand{\mean}[1]{\langle{#1}\rangle}

\newcommand{\ea}{\emph{et al.}}

\newcommand{\Omn}[1]{\emph{O}^{\chi}_{#1}}
\begin{document}

\title{Phase estimation for thermal Gaussian states}

\author{M. Aspachs}
\affiliation{Grup de F\'{\i}sica Te\`{o}rica, Universitat
Aut\`{o}noma de Barcelona, 08193 Bellaterra (Barcelona), Spain}
\author{J. Calsamiglia}
\affiliation{Grup de F\'{\i}sica Te\`{o}rica, Universitat
Aut\`{o}noma de Barcelona, 08193 Bellaterra (Barcelona), Spain}
\author{R.~Mu\~noz-Tapia}
\affiliation{Grup de F\'{\i}sica Te\`{o}rica, Universitat
Aut\`{o}noma de Barcelona, 08193 Bellaterra (Barcelona), Spain}
\author{E. Bagan}
\affiliation{Grup de F\'{\i}sica Te\`{o}rica, Universitat
Aut\`{o}noma de Barcelona, 08193 Bellaterra (Barcelona), Spain}

\date{\today}

\begin{abstract}
We give the optimal bounds on the phase-estimation precision for mixed Gaussian states in the single-copy and many-copy regimes. Specifically, we focus on displaced thermal and squeezed thermal states. We find that while for displaced thermal states an increase in temperature reduces the estimation fidelity, for squeezed thermal states a larger temperature can enhance the estimation fidelity.
The many-copy optimal bounds are compared with the minimum variance achieved by three important single-shot measurement strategies.
We show that the single-copy canonical phase measurement does not always attain the optimal bounds in the many-copy scenario.  Adaptive homodyning schemes do attain the bounds for displaced thermal states, but for squeezed states they yield fidelities that are insensitive to temperature variations and are, therefore, sub-optimal.
Finally, we find that heterodyne measurements perform very poorly for pure states but can attain the optimal bounds for sufficiently mixed states. We apply our results to investigate the influence of losses in an optical metrology experiment. In the presence of losses squeezed states cease to provide Heisenberg limited precision and their performance is close to that of coherent states with the same mean photon number.
\end{abstract}
\pacs{42.50.Dv, 03.67.Hk, 03.67.-a}

\maketitle
\section{Introduction}\label{intro}

Since the early work of Dirac \cite{dirac_quantum_1927} the problem of measuring the phase imprinted on a state of light has been a subject of  great debate.	
The problem was  to find the quantum counterpart to the classical phase observable. The observable should be the conjugate variable of the number operator $\hat n$, and the corresponding uncertainty relation could be derived. From a quantum measurement point of view the lack of such a self-adjoint observable was overcome with a different approach based on quantum estimation theory \cite{helstrom_quantum_1976, holevo_probabilistic_1982,paris_quantum_2004}, which gives a precise and operational meaning to ``measuring'' the phase (without resorting to the notion of observable) and provides reasonable uncertainty relations.
On top of this fundamental motivation, phase estimation is
at the heart of many quantum metrology \cite{giovannetti_quantum_2006} applications, such as improvement of frequency standards \cite{huelga_improvement_1997}, gravitational-wave detectors \cite{caves_quantum-mechanical_1981,goda_quantum-enhanced_2008},
and clock synchronization \cite{jozsa_quantum_2000, de_burgh_quantum_2005},
and is strongly related to quantum computation \cite{cleve_quantum_1998} and quantum cryptography \cite{gisin_quantum_2002}.
It is therefore an essential task to compute the efficiency and ultimate bounds on the precision of phase-estimation.
Optical implementations of quantum metrology applications are within reach of current technology (see, for example, recent experimental  achievements \cite{mitchell_super-resolving_2004,
higgins_entanglementfree_2007, vahlbruch_observation_2008, takeno_observation_2007, mccormick_strong_2007}).

Here we study optimal protocols to estimate the phase encoded in states of light.
In particular, we  will deal with two families of continuous variable (CV) states that are described by a Gaussian characteristic function: displaced thermal states and squeezed thermal sates.  Gaussian states are mathematically easy to handle and very relevant experimentally because they provide a very good description of the states available in the laboratories: laser fields manipulated with passive and active linear optical elements.
Although there is some previous work in the case of pure states (see \cite{monras_optimal_2006} and references therein), very little is known about estimation in mixed states, which are present in realistic unavoidably noisy scenarios.
This is the issue we tackle in this work.

In the next section we  present the general framework, introduce our figures of merit, and show that the maximum value of the average fidelity can be achieved by the single-seed covariant generalized measurement. In Sec. \ref{sec:coh} we apply the techniques introduced in Sec. \ref{sec:fw} to estimate the phase encoded in a single copy of a displaced thermal state (also known as coherent thermal states). In Sec. \ref{sec:sq}  we proceed by analyzing the problem for  squeezed thermal states of a specific class:  those that emerge when a squeezed vacuum state is sent through a lossy channel.
Contrary to intuition and in contrast with the case of coherent thermal states, we find that for squeezed thermal states the estimation precision improves with the temperature of the encoding state.
In Sec.~\ref{sec:mc} we compute the maximum fidelity when many copies of the state are available and study the performance of three particular single-shot strategies that only use individual measurements on each copy. Finally in Sec. \ref{sec:fe} we use our results to evaluate the effect of losses in metrology experiments.
We end with the main conclusions of our work.

\section{General framework}\label{sec:fw}
In this section we introduce the general problem of estimating a phase. In our setting a system evolves under a unitary transformation
\begin{equation}\label{eq:cod}
\rho(\phi)=U(\phi)\rho U(\phi)^{\dag},
\end{equation}
where $\rho$ is a general Gaussian state and $\displaystyle{U}(\phi)$
is the unitary operator
$\displaystyle{U}(\phi)=\ex{i\phi\hat{n}}$, with $\phi\in[0,2\pi)$ and $\hat{n}={a}^{\dag}{a}$ is the number operator, and ${a}^{\dag}$ and ${a}$ are the creation and annihilation operators that satisfy the bosonic commutation relations $\left[a,{a}^{\dag}\right]=1$ and $\left[{a},{a}\right]=\left[{a}^{\dag},{a}^{\dag}\right]=0$.
Although our focus here is on phase-shifted Gaussian states, the results in this section can be applied also to non-Gaussian states, in particular also to finite dimensional states, i.e. generic qudits [see however the comment after equation \eqref{eq:boundfid}].

Information on the phase is obtained through a measurement specified  by a positive operator-valued measure (POVM), i.e. by a set of operators $\{O^{\chi}\}$ that are positive semi-definite and add up to the identity. The outcome probabilities depend on the measured state through Born's rule $p(\chi|\rho)=\tr (O^{\chi}\rho)$. Estimation is thus inherently not perfect, and we need a figure of
merit to quantify how close is our guess $\phi_{\chi}$, based on outcome $\chi$, to
the real value of $\phi$. We will take the usual function
\begin{equation}
f_{l}(\phi,\phi_{\chi})=\cos[ l (\phi-\phi_{\chi})],
\end{equation}
where $ l =1$ for displaced states and $ l =2$ for squeezed
states. The factor of 2 in the second case takes into account the
symmetry of squeezed states under a phase-shift of $\pi$, due to the their double-photon structure  (see below). We will loosely refer to  $f_{l}$ as the estimation fidelity.
The corresponding average fidelity is given by
\begin{equation}
\mathcal{F}_{ l }=\sum_{\chi}{{\int_0^{2\pi}
{\frac{d\phi}{2\pi}}
f_{ l }(\phi,\phi_{\chi})\tr[\rho(\phi)\emph{O}^\chi]}}.
\label{eq:gfid}
\end{equation}

Next we proof that the maximum average fidelity can be
attained with a covariant POVM and reads as
\begin{equation}\label{eq:ofc}
\mathcal{F}_ l =\sum_{n=0}^{\infty}|\rho_{n,n+ l }|.
\end{equation}
In fact we can derive the maximum average
fidelity for a rather general family of merit functions:
$f(\phi,\phi_{\chi})=\sum_l a_{l} f_l(\phi,\phi_{\chi})=\sum_l a_l\cos
[l(\phi-\phi_{\chi})]$, with $a_l\geq 0$. This, of course, includes the above choices as well as most of the commonly used figures of merit.
To this end  we write the input state and POVM elements in the Fock basis,
\begin{equation}
\rho(\phi)=
\sum\rho_{n,
n^{\prime}}\ex{i\phi(n-n^{\prime})}\ket{n}\!\bra{n^{\prime}},
\end{equation}
$\Omn{n',n}=\bra{n'}O^\chi\ket{n}$, and develop the expression for the average fidelity to arrive to the following upper bound
\begin{align}
\mathcal{F}= &\sum_{\chi,n,n'}\int \frac{d\phi}{2\pi}
f(\phi,\phi_{\chi})
\ex{i\phi(n-n')}\rho_{n,n'}\Omn{n',n}\nonumber\\
=&\sum_{\chi,n,n',l}\int \frac{d\phi}{2\pi}  a_l \cos
[l (\phi-\phi_{\chi})]
\ex{i\phi(n-n')}\rho_{n,n'}\Omn{n',n}\nonumber\\
=& \Re \sum_{\chi,n,n',l}a_l\int \frac{d\phi}{2\pi}
\ex{i\phi(n-n'+l)}\ex{-i\phi_{\chi}l}
\rho_{n,n'}\Omn{n',n}  \nonumber\\
=&\Re\sum_{\chi,n,n',l}a_l \delta_{n',n+l}\ex{-i\phi_{\chi}l}\rho_{n,n'}\Omn{n',n}\nonumber \\
\leq &\sum_{l,n,\chi}a_{l}|\rho_{n,n+l}| |\Omn{n+l,n}|.
\label{eq:boundpr}
\end{align}
The inequality is saturated if and only if the following relation between the phases $\theta^\chi_{m,n}=\arg\Omn{m,n}$ and $\psi_{m,n}=\arg\rho_{m,n}$ holds:
\begin{equation}
\theta^\chi_{n,m}=\psi_{n,m}+\phi_{\chi}(n-m)  \; \; \forall n,m.
\label{eq:phasecond}
\end{equation}
Now, from the positivity of $\emph{O}^\chi$ and the fact  that the geometric mean is bounded from above by the arithmetic mean, it follows $ |\Omn{n,m}|\leq \sqrt{\Omn{n,n}\Omn{m,m}}\leq 1/2(\Omn{n,n}+\Omn{m,m})$, which together with the POVM completeness relation
$\sum_{\chi}\Omn{n,m}=\delta_{n,m}$ leads to
\begin{equation}
\mathcal{F}\leq \sum_{l}a_{l}\sum_{n}|\rho_{n,n+l}|.
\label{eq:boundfid}
\end{equation}
This bound can be attained iff $|\Omn{m,n}|=|\Omn{m,m}|=|\Omn{n,n}|$ for all $m,n$, and $\chi$.
This is precisely  satisfied by the well-known (canonical) \emph{ phase-measurement} \cite{holevo_probabilistic_1982}. This is given by a continuous POVM [labeled by $\theta\in[0,2\pi)$ instead of a discrete index $\chi$] with elements
$\emph{O}^\theta=1/(2\pi) \ketbrad{\theta}$ where $\ket{\theta}=\sum_{n}\ex{i \theta n}\ket{n}$.
Condition  \eqref{eq:phasecond} on the phases is automatically met if the density matrix $\rho$ has positive entries in the Fock basis, i.e., $\psi_{n,m}=0$, and the phase
$\theta$  is guessed whenever the measurement outcome $\emph{O}^\theta$ occurs.
If $\rho$ does have non-zero phases of the form $\psi_{n,m}=\zeta(n)-\zeta(m)$, i.e. they can be removed by a unitary transformation $U=\sum_{n}\ex{i\zeta(n)}\ketbrad{n}$, the phase measurement given above needs to be corrected by the corresponding unitary,  i.e., $\ket{\theta}=\sum_{n}\ex{i [n \theta +\zeta(n)]}\ket{n}$.
For more general phases of $\rho_{n,m}$,  the bound holds  if condition \eqref{eq:phasecond} does not conflict with the positivity condition on the POVM elements. For any dimension $d>2$ one can easily find examples of seed states $\rho(0)$ with arbitrary phases for which the bound cannot be attained. However, for the Gaussian states under study here, phases can always be unitarily cancelled, and therefore \eqref{eq:ofc} does indeed give the maximum fidelity.
In fact, for the figures of merit that we use in this paper, and in general for those that only have a single Fourier component (only one non-zero coefficient~$a_{l}$),
the bound can always be attained  since any arbitrary phase $\psi_{n,n+l}$ can be unitarily reabsorbed by the POVM, i.e., $\ket{\theta}=\sum_{n}\ex{i [n \theta +\xi(n)]}\ket{n}$, where the phases satisfy the recursion relation $\xi(n+l)=\xi(n)+\psi_{n,n+l}$, with $\xi(n)=0$ for~$n<l$.

Note that the phase measurement is optimal not only for a large family of input states but also for a very wide class of figures of merit. A remarkable exception occurs for the state estimation fidelity $f(\rho_{\chi},\rho_{\phi})=(\tr |\sqrt{\rho_{\chi}}\sqrt{\rho_{\phi}}|)^2$, where $\rho_{\chi}$ is the state ``guessed'' on measurement outcome $\chi$.  This optimal guess state is not necessarily one of the possible input states \cite{calsamiglia_phase-covariant_2008} and the figure of merit cannot always be written as a Fourier series with positive coefficients.  Therefore the above derivation does not hold. In fact, in state estimation of phase-covariant states there are examples where the phase measurement is known to be sub-optimal \cite{calsamiglia_phase-covariant_2008}.

Equation \eqref{eq:boundfid}
provides the ultimate bounds in phase estimation; however the phase measurement saturating this bound may be very difficult (if not impossible)   to implement experimentally in a single-shot (or single copy) measurement \cite{wiseman_adaptive_1995}. However, as we will see in section \ref{sec:mc}, when a large number of copies are available there exist known measurement schemes that attain the optimal bounds in some particular cases.

The remaining of this section is devoted to  Gaussian states of light.
The Gaussian states are those that are fully characterized by the first and second moments of the field quadratures, i.e. by the displacement vector ${d}=\tr({R}\rho)$ and the covariance matrix $\Gamma_{kl}=\tr(\rho \{R_{k}-d_{k},R_{l}-d_{l}\}_{+})$, where $R_{1}=1/\sqrt{2}(a+a\dg)$ and $R_{2}=i/\sqrt{2}(a\dg-a)$ are  (conjugated) quadratures.

An equivalent, perhaps more operational, definition can be given in terms of the action
of the squeezing operator ${S}(r)=\exp[\frac{r}{2}( a^2- {a^\dagger}^2)]$, and the  displacement operator $
D(\alpha)=\ex{\alpha a\dg-\alpha^* a}$  on a thermal state
\be
\rho_{\beta}=(1-\ex{-\beta}) \sum_n \ex{-\beta n} \ketbrad{n},
\ee
namely,
\begin{equation}
\rho_{\beta,\alpha,r}(\phi)= U(\phi)D(\alpha) {S}(r)
\rho_{\beta}{S}(r)\dg D(\alpha)\dg U(\phi)\dg.
\end{equation}
For a general Gaussian state the mean photon number can be easily seen to be
\be\label{eq:mphnum}
\ve{n}=|\alpha|^2 +n_{\beta}+(2n_{\beta}+1)\sinh^2 r,
\ee
where $n_{\beta}=(\ex{\beta}-1)^{-1}$ is the thermal mean photon number.
The covariance matrix of a thermal state is simply $\Gamma_{\beta}=\gamma_{\beta} \id$, with $\gamma_{\beta}=\tanh^{-1}
(\beta/2)=2 n_{\beta}+1$.
The squeezing operator $S(r)$ decreases the fluctuations in one quadrature and increases them by the same factor in the second quadrature:
\begin{equation}
\label{eq:covmaa}
\Gamma_{r,\beta}=\!(2{n}_{\beta}+1)\!\!
                 \begin{pmatrix}
                   \ex{-2r} & 0 \\
                   0 & \ex{2r}
                 \end{pmatrix} .
\end{equation}
The displacement $D(\alpha)$ does not change the second moments, instead it only induces a displacement ${ d}=\sqrt{2}(\mathrm{Re} \alpha,\mathrm{Im} \alpha)^T$.
Finally the phase operator $U(\phi)$ produces a rotation in phase space which induces the  corresponding transformation on the displacement vector and covariance matrix : ${ d}'=O_{\phi} { d}$
and $\Gamma'=O_{\phi}\Gamma O_{\phi}^T$ with
\be
\label{eq:rot}
O_{\phi}=\left(\begin{array}{cc}\cos\phi &
-\sin\phi \\\sin\phi &\cos\phi \end{array}\right).
\ee

We will next concentrate on two classes of Gaussian mixed states: displaced thermal states, also called coherent thermal states, and squeezed thermal states.

\section{Coherent Thermal States}\label{sec:coh}

Coherent states, defined as
\be
\ket{\alpha}=D(\alpha)\ket{0}=\ex{-\frac{|\alpha|^2}{2}}\sum_{n=0}^\infty
\frac{\alpha^n}{\sqrt{n!}}\ket{n},
\ee
are very relevant in CV  implementations of quantum information protocols since they provide a very good description of the states produced by a laser: they are states with a well-defined amplitude and phase, and with minimal fluctuations in both quadratures (see however \cite{mlmer_optical_1997} and \cite{vanEnkLaser} for some caveats on this description of the output of a laser). A displaced thermal state  $\rho_{\beta,\alpha}=D(\alpha)\rho_{\beta}D(\alpha)\dg$ is also well represented by an amplitude and a phase but  fluctuations are larger.

\begin{figure}[!htp]
\includegraphics[width=3.0in]{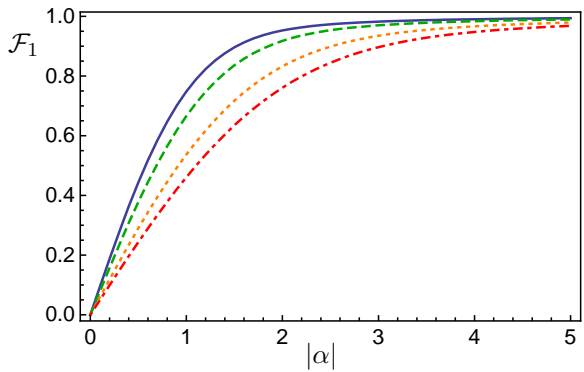}
\caption{(Color
 online) Average fidelity for coherent thermal states for different ${n_{\beta}}$: ${n}_{\beta}=0.1$ (solid), ${n}_{\beta}=0.5$
(dashed), ${n}_{\beta}=1.5$ (dotted), ${n}_{\beta}=2.5$ (dash-dotted).
\label{fig:fthch}}
\end{figure}

In order to calculate the maximum fidelity \eqref{eq:ofc} it is convenient to write the thermal state in the $P$-function representation
\begin{equation}
\rho_{\beta}=\frac{1}{\pi {n}_{\beta}}\int{d^{2}\alpha'
\ex{-\frac{|\alpha'|^{2}}{{n}_{\beta}}}\ketbra{\alpha'}{\alpha'}}.
\end{equation}
One can check that this is indeed a thermal state just by
computing the matrix elements in the Fock basis
\begin{align}
\bra{n}\rho_{\beta}\ket{m}&=\frac{1}{\pi{n}_{\beta}}\int d^{2}\alpha'
\ex{-\frac{|\alpha'|^{2}}{{n}_{\beta}}}\braket{n}{\alpha'}\braket{\alpha'}{m}=\nonumber\\
&=\delta_{nm}\frac{{n}_{\beta}^{n}}{(1+{n}_{\beta})^{n+1}}.
\end{align}
In the Fock basis the coherent thermal state reads as
\begin{align}
\label{eq:rhocohtherm}
&\rho_{\beta,\alpha}=\frac{1}{\pi{n}_{\beta}}\int{d^{2}\alpha'
\ex{-\frac{|\alpha'|^{2}}{{n}_{\beta}}}\ketbrad{\alpha'+\alpha}}=\\
&=\sum_{k,l}{\int
{\frac{d^{2}\alpha'}{\pi{n}_{\beta}}\ex{-\frac{|\alpha'|^{2}}{{n}_{\beta}}
}\ex{-|\alpha'+\alpha|^{2}}\frac{(\alpha'+\alpha)^{k}(\alpha'+\alpha)^{*l}}
{\sqrt{k!}\sqrt{l!}}\ketbra{k}{l}}}. \nonumber
\end{align}
Equation (\ref{eq:ofc}), for $l=1$, gives the maximum fidelity, which reads:
\begin{equation}
\mathcal{F}_1=\sum_{k}\int\frac{d^{2}\alpha'}{\pi{n}_{\beta}}\ex{-\frac{|\alpha'|^{2}}{{n}_{\beta}}}
\ex{-|\alpha'+\alpha|^{2}}\frac{|\alpha'+\alpha|^{2k}(\alpha'+\alpha)^*}{k!\sqrt{k+1}}.
\end{equation}
At this point we use the following integral representation,
\begin{equation}\label{eq:intform}
\frac{1}{\sqrt{k+1}}=\frac{1}{\sqrt{\pi}}\int_{0}^\infty{\frac{dt}{\sqrt{t}}\ex{-t(k+1)}}
\end{equation}
that allows us to evaluate the sum. The integral over $\alpha'$ can then be easily evaluated:
the imaginary part is trivially zero and the real part is a Gaussian integral that is equal to
\begin{equation}\label{eq:fc}
\mathcal{F}_{1}=\frac{|\alpha|}{\sqrt{\pi}}\int_{0}^{1/(1+n_{\beta})}{dy\frac{\ex{-y |\alpha|^{2}}}{\sqrt{-\ln[1-y/(1-n_{\beta} y)]}}},
\end{equation}
after the change in variable $y=(1-\ex{-t})/[1+n_{\beta}(1-\ex{-t})]$.
This expression can be evaluated numerically with arbitrary precision for different thermal photon numbers $n_{\beta}$. Figure \ref{fig:fthch} shows the average
fidelity as a function of the field amplitude for different temperatures. As one could expect, higher temperatures increase the fluctuations without changing the field amplitudes and therefore decrease the precision of the estimation.

The asymptotic behavior of the fidelity at large field amplitudes can be computed analytically from (\ref{eq:fc}) by expanding the non-exponential factor of the integrand up to second order,
\begin{equation}
\mathcal{F}_1\simeq 1-\frac{2{n}_{\beta}+1}{8 n_{\alpha}} ,
\label{eq:fidcohas}
\end{equation}
where $n_{\alpha}=|\alpha|^2$ is the mean photon number of a coherent state $\ket{\alpha}$, or the contribution of  the displacement $D(\alpha)$ to the mean photon number of a coherent thermal state $\bar{n}={n}_{\beta}+\left|\alpha\right|^{2}$.
In the limit of zero temperature, ${n}_{\beta}=0$,
we recover the result for pure states $\mathcal{F}_1\simeq1-{1}/{(8\bar{n})}$ derived  in \cite{bagan_phase_2008}.

We next turn to the opposite regime.  In  order to compute the fidelity for low-field amplitudes ($\alpha\to 0$), we write the thermal state in the Fock basis and expand $D(\alpha)$ up to first order in $\alpha$,  $D(\alpha)\approx \id-\alpha(a-a\dg)$. With this one can easily obtain a closed expression for the fidelity up to first order in $|\alpha|$.  For simplicity here we only give its asymptotic behavior at high and low temperatures:
\be
\mathcal{F}_1=
\left\{\begin{array}{lcc}
\displaystyle \sqrt{\frac{\pi}{2}}\sqrt{\frac{n_{\alpha}}{2n_{\beta}+1}}
&\mbox{  for  } \; \; & n_{\beta}\gg 1  \\
\\
\sqrt{n_{\alpha}} [1-(2-\sqrt{2})n_{\beta}]
&\mbox{  for  } \; \;& n_{\beta}\ll 1 .
\end{array}\right.
\ee

\section{Squeezed Thermal States}\label{sec:sq}
Squeezed vacuum states are defined as
\be
\ket{r}=\!S(r)\ket{0}\!=\!(1-\lambda^2)^{1/4}
\sum_{n=0}^\infty\!\left(\!-\frac{\lambda}{2}\right)^n\!\! \frac{\sqrt{(2n)!}}{n!}\!\ket{2 n},
\label{eq:sqvac}
\ee
where $\lambda$ is related to the squeezing parameter $r$ and the mean photon number $n_{r}=\sinh^2r$ by $\lambda=\tanh{r}=\sqrt{n_{r}/(n_{r}+1)}$. Squeezed states are those for which the uncertainty (or fluctuations) on a given quadrature is reduced below the standard quantum limit (the quantum fluctuations of vacuum) at the expense of increasing the uncertainty in the conjugate quadrature.  One can therefore expect an enhanced performance of squeezed states in phase estimation or in other high-precision applications \cite{walls_squeezed_1983,treps_quantum_2003,goda_quantum-enhanced_2008}. This states exhibit very non-classical features, such as sub-Poissonian statistics or the ability to generate entanglement \cite{ou_realization_1992,kim_entanglement_2002}, which allows for a variety of applications in quantum information theory \cite{braunstein_quantum_2005}. Squeezed states have been successfully produced in the laboratories for some time now \cite{slusher_observation_1985,wu_generation_1986} reaching squeezing of up to  $10$ dB (corresponding to $r\lesssim 1.15$ or $n_{r} \lesssim 2$) in current experiments  \cite{vahlbruch_observation_2008,takeno_observation_2007,mccormick_strong_2007}.
Squeezed thermal states are those obtained by squeezing an initial thermal state (instead of the vacuum),
\be
\rho_{\beta,r}=S(r)\rho_{\beta}S(r)\dg.
\ee
Therefore one expects quantum features to be less pronounced. At high enough temperatures  [for $\ex{-2 r}(2 n_{\beta}+1)>1$] the state will even become classical \cite{marian_squeezed_1993}, i.e. a mixture of coherent states,  loosing its entanglement potential \cite{asboth_computable_2005} and other quantum features.

In order to obtain the maximum fidelity \eqref{eq:ofc} for squeezed thermal state we only need to sum the outer diagonal elements $|\rho_{n,n+2}|$. A closed form expression for  the matrix elements $\bra{m}S(r)\ket{n}$ can be found in \cite{satyanarayana_generalized_1985} for arbitrary Fock states, but it is highly non-trivial and turns the evaluation of the fidelity into a very hard computation.
Let us hence consider a much more tractable, but still very relevant, family of squeezed thermal states.

Most sources of noise in quantum optical experiments (e.g., mode
mismatch, misalignment, absorption of optical elements, and non-unit
detector efficiencies) can be cast in terms of linear losses. The family that we will study here consists of the mixed states that arise
when a squeezed vacuum undergoes losses during the course
of an experiment. Since linear losses can be modeled
by a  beam splitter (BS) of transmittance $T$, this family has the following simple characterization:
\be
\rho_{\beta,r}(\phi)=\tr_{b}\left(B_{\theta}U_{a}(\phi)\ket{r_{0} 0}_{ab}
\!\!\bra{r_{0} 0}U_{a}(\phi)\dg
B_{\theta}^{\dag}\right),
\nonumber
\label{eq:sqthermal}
\ee
where ${B}_{\theta}=\exp[\theta({a}^{\dag}{b}-{a}{b}^{\dag})]$, with $T=1-R=\cos^2
{\theta}$, is the BS transformation  that acts on both the system (mode $a$) and the environment (mode $b$, with $[b,b\dg]=1$), which is initially in the vacuum state. Notice that the effect of losses commutes with the phase operation and, therefore, the states under consideration are of form \eqref{eq:cod}, i.e.,
$\rho(\phi)=U(\phi)\rho U(\phi)\dg$. In what follows we find how the input squeezing parameter $r_{0}$, and the transmittance of the channel, $T$, are related to the inverse temperature,~$\beta$, and squeezing parameter $r$ of the output state.

The covariance matrix associated with the two-mode initial state,
$\ket{r_{0}}_{a}\otimes\ket{0}_{b}$, is given by
$\Gamma_{\mathrm{in}}=\Gamma_{r_{0},0}\oplus\id$. The action of the BS transforms
the quadratures according to the symplectic operation,
\begin{equation}
V_{\theta}=
\begin{pmatrix}
\cos\theta & 0 & \sin\theta & 0 \\
0 & \cos\theta & 0 & \sin\theta \\
-\sin\theta & 0 & \cos\theta & 0 \\
0 & -\sin\theta & 0 & \cos\theta
\end{pmatrix} ,
\end{equation}
leading to the two-mode covariance matrix
\begin{widetext}
\begin{equation}\label{eq:gammaout}
\Gamma_{\mathrm{out}}={V}_{\theta}^{T}\Gamma_{\mathrm{in}}{V}_{\theta}=\begin{pmatrix}
  \ex{-2r_{0}}T+R & 0 & (\ex{-2r_{0}}-1)\sqrt{TR} & 0 \\
  0 & \ex{2r_{0}}T+R & 0 & (\ex{2r_{0}}-1)\sqrt{TR} \\
  (\ex{-2r_{0}}-1)\sqrt{TR} & 0 & \ex{-2r_{0}}R+T & 0\\
  0 & (\ex{2r_{0}}-1)\sqrt{TR} & 0 & \ex{2r_{0}}R+T
\end{pmatrix}.
\end{equation}
\end{widetext}
If we consider only mode $a$ we obtain
\begin{equation}\label{eq:mcovbs}
\Gamma_{\mathrm{out}}^{a}=
\begin{pmatrix}
  \ex{-2r_{0}}T+R & 0  \\
  0 & \ex{2r_{0}}T+R
\end{pmatrix}.
\end{equation}\\
That is, the effect of losses turns the initial squeezed vacuum $\ket{r_{0}}$ into a squeezed thermal state $\rho_{\beta,r}$. Equating the covariance matrix \eqref{eq:mcovbs} to its general form for squeezed thermal states \eqref{eq:covmaa} we find
the following equations relating input and output parameters,
\begin{align}\label{eq:par}
&2{n}_{\beta}+1=\tanh^{-1}\frac{\beta}{2}=\sqrt{T^2+2 TR \cosh{2r_{0}}+R^2},\nonumber\\
&\mathrm{e}^{2 r}= \sqrt{\frac{\mathrm{e}^{
2r_{0}}T+R}{\mathrm{e}^{ -2r_{0}}T+R}}.
\end{align}
Note that equations \eqref{eq:par} do not have solution for all values of $n _{\beta}$ and $r$. That is,  not all squeezed thermal states can be viewed as a squeezed vacuum that has degraded through a lossy channel.

Now we can write the density matrix in the Fock basis for this family of squeezed thermal states by sending a squeezed vacuum $\ket{r_{0}}$ \eqref{eq:sqvac} through a  BS \eqref{eq:sqthermal} and tracing out mode  $b$,
\begin{align}\label{eq:encodest2}
&\rho_{\beta,r}(\phi)=\sqrt{1-\lambda_{0}^{2}}\sum_{n,m=0}^{\infty}\left(-\frac{\lambda_{0}}{2}\right)^{n+m}\ex{2i(n-m)\phi}\\
&\times \frac{\sqrt{(2n)!}\sqrt{2m!}}{n! m!}\sum_{k}\!\sqrt{p^{2n}_T(k)}\sqrt{p^{2m}_T(k)}
\ket{2n-k}\!\bra{2m-k},\nonumber
\end{align}
with $\lambda_{0}=\tanh r_{0}$ and with the binomial distributions $p^{2n}_T(k)$ and $p^{2m}_T(k)$ given by
\begin{equation}\label{eq:binonomial}
 p^{N}_T(k)= \frac{N!}{(N-k)!k!}R^k T^{N-k}.
\end{equation}
Using  (\ref{eq:ofc}) the maximum fidelity can be
written as:
\begin{align}
\mathcal{F}_2&=\sqrt{1-\lambda_{0}^2}\sum_{n,k}
\left(\frac{\lambda_{0}}{2}\right)^{2n+1}\nonumber\\
\times&\frac{(2n)!(2n+2)!\;R^{k}\;T^{(2n+1-k)}}{n!(n+1)!k!(2n-k)!
\sqrt{2n+2-k}\sqrt{2n+1-k}}.
\label{eq:fidTR}
\end{align}
At this point, we use the integral representation
(\ref{eq:intform}) for both ${1}/{\sqrt{2n+2-k}}$ and
${1}/{\sqrt{2n+1-k}}$, which reads as
\begin{align}
&\frac{1}{\sqrt{2n+2-k}}\frac{1}{\sqrt{2n+1-k}}=\nonumber\\
&\frac{1}{\pi}\int_{0}^{\infty}
{\frac{dx}{\sqrt{x}}\ex{-x(2n+2-k)}}\int_{0}^{\infty}{\frac{dy}{\sqrt{y}}\ex{-y(2n+1-k)}}.
\end{align}
This enables us to perform the sum over $n$ and $k$ explicitly.
After the change of variables $x=ut$ and
$y=(1-u)t$ one easily identifies the integral representation of the
Bessel function
$$I_{0}(t)=\frac{1}{\pi}\int_{0}^{1}{\frac{du}{\sqrt{u(1-u)}}\ex{t(1-2u)}}$$
and the fidelity can be cast in the compact form
\begin{equation}\label{eq:fidsth}
\mathcal{F}_2=\lambda_0\sqrt{1-\lambda_0^2} \int_{0}^\infty\!\!\!\!\!\! \frac{dt\;
\mathrm{e}^{-\frac{3}{2}t}
I_{0}(t/2)\;T}{\{1-\lambda_{0}^2[R+T\mathrm{e}^{-t}]^2\}^{3/2}},
\end{equation}

\begin{figure}[!htp]
\includegraphics[width=3.2in]{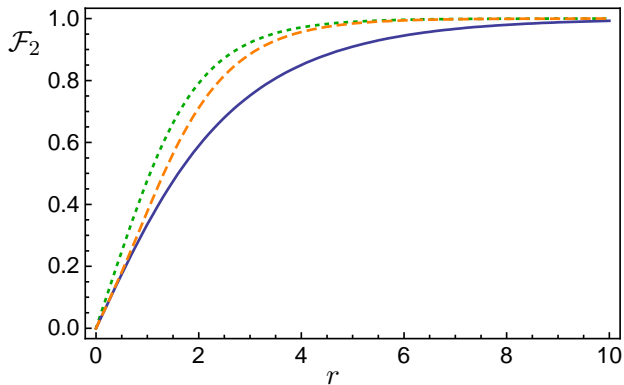}
\caption{(Color
 online) Fidelity for pure squeezed state $\ket{r}$ (solid) and for states with the same squeezing parameter emerging from a 50/50 BS (dashed). We also show the fidelity for a pure squeezed state $\ket{\tilde r}$ with the same energy as the  former thermal states, i.e. $\mean{n}=\sinh^2 \tilde r
=1/2\sinh^2 2r$ (dotted line).}\label{fig:fths}
\end{figure}

From Equations \eqref{eq:par}  relating  $r$ and $n_{\beta}$ with $r_{0}$ and $T$
it is straightforward to calculate the fidelity for states that have the same squeezing parameter $r$ but different temperature.
In Figure  \ref{fig:fths} we show the fidelity as a function of $r$ for a pure state, $\ket{r}$, 
and for a mixed state.
The latter is taken to be the state that results from sending  $\ket{r_{0}=2 r}$ through a 50:50 BS ($T=1/2$), which indeed has a squeezing parameter equal to $r$. We see that the mixed states (non-zero temperature) allow for  better phase estimation than pure states (zero temperature). This behavior is a bit puzzling at first sight and opposite to what we found for coherent states. It is true that noise \emph{per se} increases the uncertainty in any estimated quantity. However, increasing the temperature has a second effect that goes in the opposite direction. When the squeezing operator acts on a thermal state the number of photons increases in a non-linear fashion,
\be
\ve{n}=n_{\beta}+(2n_{\beta}+1)\sinh^2 r.
\ee
This boosted increment gives rise to an improvement on the phase sensitivity that outweighs
the adverse effects of the temperature. In the case of coherent states,
the action of displacement adds the same amount of energy independently of the temperature, and hence the detrimental effect of thermal noise is not counterbalanced by the energy increase.
Figure \ref{fig:fths} also shows (dotted line) the fidelity of a pure squeezed $\ket{\tilde r}$ state with the same mean energy of the squeezed thermal state, i.e., $n_{\tilde r}=\sinh^2 \tilde r=1/2 \bra{2r}\hat{n}\ket{2r}=1/2 \sinh^2 (2r)$, where we have used that the BS is balanced and half of the photons are lost. We find that, given a fixed mean energy, it is  clearly more advantageous to increase the squeezing parameter rather than the temperature.

The fidelity behavior in the asymptotic limit when the
squeezing parameter is large can be computed analytically.
With the change in variables $t=-\ln w$, Eq.(\ref{eq:fidsth}) becomes
%
\begin{equation}
\label{e-ebc1}
\mathcal{F}_2=
\lambda_0\sqrt{1-\lambda_0^2}\;T\int_0^1\frac{dw \, \sqrt w\, I_0(\frac{1}{2}\ln w)}{[1-\lambda_0^2(R+T w) ^2]^{3/2}}.
\end{equation}
%
Note that in the limit we are interested in, $\lambda_0$ approaches unity and the dominant contribution to the fidelity comes from the region~$w\approx1$. It is, therefore, convenient to separate this contribution by writing the numerator of~(\ref{e-ebc1}) as $1-[1-\sqrt w I_0(\frac{1}{2}\ln w)]$ and integrating the two terms separately. The integral corresponding to the first term (unity) is straightforward and contributes to both leading and subleading orders. The term in square brackets goes as $1-w$ for $w\approx1$ and is subleading; thus we can safely put $\lambda_0=1$ in the corresponding integral, as the neglected terms will give even higher order contributions to~$\mathcal{F}_{2}$. The result can be cast~as
\begin{equation}
\mathcal{F}_2=1-\frac{\xi(T)}{ \sqrt{n_{0}}}  ,
\end{equation}
where
\be
\xi(T)= \frac{R}{\sqrt{T(1+R)}}+
T\int_{0}^1\!dw\frac{1-\sqrt{w} I_0({\frac{1}{2}}\ln w)}{[1-(R+T w)^2]^{3/2}} .
\ee

The remaining integral can be computed numerically to arbitrary accuracy
for any
value of  the channel transmittance~$T$. For $T=1$ (pure squeezed state) we obtain~$\xi(1)=0.55$, in agreement with~\cite{bagan_phase_2008}. We further note that, as expected, $\xi$ is a decreasing function of $T$.
Thus, e.g., $\xi(1/2)=1.20$ and
$\xi(1/3)=1.62$.  By  interpolating $\xi(T)$ between the two regimes where it can be analytically computed ($T\to 1$ and $T\to 0$) one can check that
to an accuracy of more than $98\%$, one has
\begin{equation}
\xi(T)\approx \frac{R}{\sqrt{T(1+R)}}+\frac{c_1}{\sqrt{2T}}+c_2 T ,
\end{equation}
where $c_1=0.54$ and $c_2=0.17$.

Let us now consider the opposite limit of low energy, which is relevant because the amount of squeezing available in laboratories is often quite limited. For $\lambda_{0}\ll 1$ we can take $\lambda_{0}\approx\sqrt{n_0}$ and retain only the first term ($n=0$) in the sum  (\ref{eq:fidTR}) to get
\begin{equation}
\mathcal{F}_2\simeq\sqrt{\frac{n_0}{2}}T,
\end{equation}
or $\mathcal{F}_2\approx T r_{0}/{\sqrt{2}}\approx r/\sqrt{2}$ in terms of the squeezing parameter. This explains the linear behavior shown
 in Fig.~\ref{fig:fths} for small values of $r$.

\section{Many copies}\label{sec:mc}

The calculation of the fidelity when several copies $N$ are available
is in general a hard task. The case of pure states has only been solved  recently
\cite{bagan_phase_2008} by computing the matrix elements of $|\rho^{\otimes N}_{n ,n+l}|$ in a nondegenerate eigen-basis of $\hat{n}_{t}=\sum_{i=1}^N a_{i}\dg a_{i}$. The case of mixed states becomes further involved  since the basis that spans $\rho^{\otimes N}$ is degenerate with respect to $\hat{n}_{t}$.
Here we will follow a different approach to calculate the asymptotic behavior of the fidelity in the limit of large number of copies.

We first notice that in this limit the estimation becomes very accurate ($\phi-\phi_{\chi}$ will be typically small) and therefore we can relate  the estimation fidelity \eqref{eq:ofc} to the variance:  $\mathcal{F}_{1}=1-\mathrm{Var}[\phi]/2$ for coherent states, and
$\mathcal{F}_{2}=1-2\mathrm{Var}[\phi]$ for squeezed states.
The Cram\'{e}r-Rao bound \cite{cramer},  a well-known result in classical statistics,
gives a lower bound on the variance of an unbiased estimator in terms
of the Fisher information $I(\phi)$, which is a functional of the parametric family of probability distributions $p(\chi|\phi)$ from where the samples are drawn
\begin{equation}
\label{eq:fisher}
I(\phi)=\int{d\chi p(\chi|\phi)\left(\frac{\partial\ln p(\chi|\phi)}{\partial\phi}\right)^2}.
\end{equation}
For a large number of samples $N$ and under some regularity conditions,
the Cram\'{e}r-Rao bound is attained and we have that $\mathrm{Var}^{\mathrm{opt}}[\phi]=( N I)^{-1}$.
In their seminal paper Braunstein and Caves \cite{braunstein_statistical_1994} made use of this result to proof that if $N$ copies of a state $\rho(\phi)$ are available, the optimal estimation of $\phi$ based on the most general quantum measurement is 
\be
\label{eq:brcaves}
\mathrm{Var}^{\mathrm{opt}}[\phi]= \mathrm{d}\phi^2/ (4 N \mathrm{d}s^2_{\mathrm{BU}}),
\ee
where the Bures metric $ \mathrm{d}s^2_{\mathrm{BU}}$ can be obtained from the Bures distance between two infinitesimally close states: $ \mathrm{d}s^2_{\mathrm{BU}}=1-F(\rho,\rho-\mathrm{d}\rho)$, where $F(\rho_{1},\rho_{2})=\tr\left(\sqrt{\sqrt{\rho_1}\rho_2\sqrt{\rho_2}}\right)^2$ is the quantum fidelity \cite{jozsa_fidelity_1994}. Closed expressions for this fidelity have been computed in \cite{twamley_bures_1996, scutaru_fidelity_1998, paraoanu_bures_1998}  for general Gaussian states.
Therefore, in order to obtain the maximum estimation fidelity we only need to compute $F[\rho(0),\rho(\mathrm{d\phi)}]$ for infinitesimally close Gaussian states.

For displaced thermal states, $\rho_{\beta,\alpha}$, we find
\begin{equation}\label{eq:varcoh}
\var_{\mathrm{coh}}^{\mathrm{opt}}[\phi]=\frac{1}{4N} \frac{1}{|\alpha|^2 \tanh\beta/2}=\frac{1}{4N}\frac{2 n_{\beta}+1}{n_{\alpha}}.
\end{equation}
For pure coherent states we recover the known results \cite{monras_optimal_2006} which agree with the standard quantum limit or shot noise limit ($\var[\phi]\sim n^{-1}$).
We notice here that for pure states the $N$-copy result follows straightforwardly from the single-copy case since an $N$-copy coherent state $\ket{\alpha}^{\otimes N}$ is unitarily equivalent to a large amplitude coherent state $|{\sqrt{N} \alpha}\rangle$. The unitary can be simply implemented by a linear multiport device that realizes a linear mode transformation such that $a_{0}=1/\sqrt{N}(a_{1}+a_{2}+\ldots a_{N})$.
For mixed states the  error increases linearly with the size of the quadrature fluctuations, i.e. the phase variance scales linearly with the number of thermal photons $n_{\beta}$. We also notice that although the above strict equivalence between the single-copy and many-copy regimes does not hold for mixed states, the variance at large field amplitudes  has exactly the same behavior in both regimes [compare \eqref{eq:varcoh} with \eqref{eq:fidcohas}].

For squeezed thermal states, $\rho_{\beta,r}$, we find,
\begin{align}\label{eq:varenen}
\var_{\mathrm{sq}}^{\mathrm{opt}}[\phi]&=\frac{1}{2N}\frac{1}{(1+\cosh^{-1}{\beta})\sinh^{2}{2r}}\nonumber\\
&=
\frac{1}{16 N n_{r}(n_{r}+1)}\left[1+\frac{1}{(2 n_{\beta}+1)^2}\right],
\end{align}
where we used that $\sinh^2(2r)=4{n}_r({n}_r+1)$. Again, for pure states we recover the well-known result in \cite{yurke_su2_1986, monras_optimal_2006}, which has the scaling of the so-called Heisenberg limit $\var[\phi]\sim n^{-2}$.  Here, as in the single-copy scenario, we find that even though the phase-space distribution clearly shows that  temperature increases the size of the fluctuations, a higher temperature prior to squeezing does in fact improve the estimation fidelity. Notice that this behavior appears at all temperatures and persists even for classical states, i.e., when $(2n_{\beta}+1)>\exp({2r})$ (see above).

Finally, we would like to write the minimal variance in terms of the parameters of the lossy channel described in the previous section, i.e., the squeezing parameter $r_{0}$ of the initial squeezed vacuum and the transmittance $T$ of the channel.
Using relation \eqref{eq:par} the optimal variance can be written after some algebra as
\begin{equation}
\var_{\mathrm{sq}}^{\mathrm{opt}}[\phi]=\frac{1+2 T(1-T)\sinh^2 r_{0}}{2 N  T^2 \sinh^2(2r_{0})},
\end{equation}
or in terms of the input's mean photon number $n_0=\sinh^2 r_{0}$,
\begin{equation}
\label{eq:optvarT}
\var_{\mathrm{sq}}^{\mathrm{opt}}[\phi]=\frac{1+2 T(1- T) n_0}{8 N  T^2 \;n_0(n_0+1)},
\end{equation}
which in the limits of high and low squeezing gives
\begin{eqnarray}
\label{eq:ophs}
\Var_{\mathrm{sq}}^{\mathrm{opt}}[\phi]&=&\frac{1-T}{4 N T n_0} \hspace{1cm} \mbox{for }\; n_0\gg1,\\
\Var_{\mathrm{sq}}^{\mathrm{opt}}[\phi]&=&\frac{1}{8 N T^2 n_0}
\hspace{.8cm} \mbox{for }\; n_0\ll1.
\label{eq:opls}
\end{eqnarray}
This shows that the Heisenberg limited precision cannot be attained in the presence of losses.

The results we have presented so far for the many-copy scenario are again theoretical bounds on the optimal variance based on relation \eqref{eq:brcaves}. The proof of this relation is constructive in the sense that it provides a particular single-copy measurement and an estimator that saturate the bound.  However,
this optimal measurement is given in terms of an observable called symmetric logarithmic derivative \cite{braunstein_statistical_1994}, which typically cannot be implemented experimentally. In the following subsections we study three  particular single-shot  measurements and compare their performance with the optimal bounds.

\subsection {Heterodyne measurements}

Heterodyning is one of the first and most common detection schemes for state reconstruction in CV systems for it yields direct information about the phase-space distribution (Q-function) that is enough to completely reconstruct the state.
The POVM of an ideal heterodyne measurement is given by  $\{1/\pi\ketbrad{\alpha}\}$.

The overlap between two arbitrary Gaussian states $\rho_{A}$ and $\rho_{B}$ is given by,
\begin{equation}
\label{eq:overlap}
\tr(\rho_{A}\rho_{B})=2\left[\det(\Gamma_{A}+\Gamma_{B})\right]^{-\frac{1}{2}}\ex{-  \delta^t
(\Gamma_{A}+\Gamma_{B})^{-1} \delta} ,
\end{equation}
where $\delta$ is the difference between the displacement vectors of the two states $\delta=d_{A}-d_{B}$, and $\Gamma_{A}$ and  $\Gamma_{B}$ are the covariance matrices of each state.
Therefore, a heterodyne measurement on a squeezed thermal state $\rho_{\beta,r}(\phi)$  will yield  outcome $\alpha'$, or more precisely its real and imaginary parts
$\{\alpha'_{x},\alpha'_{p}\}=\{\mathrm{Re}[\alpha'], \mathrm{Im}[\alpha']\}$, with probability,
\begin{equation}
\label{eq:hete-probability2}
 p(\alpha' |\phi)=\frac{1}{\pi}\tr [\ketbrad{\alpha'} \rho_{\beta,r}] =
\frac{  \ex{- d^t O_{\phi}\Gamma_{S}^{-1}O_{\phi}^t d}
}{\pi\sqrt{V_{+}V_{-}}},
\end{equation}
where $d=\sqrt{2}\{\alpha'_{x},\alpha'_{p}\}$, $O_{\phi}$ is defined in \eqref{eq:rot}, and
\begin{equation}
\Gamma_{S}=
                 \begin{pmatrix}
                   V_{-} & 0 \\
                   0 &V_+
                 \end{pmatrix}
\end{equation}
with $V_{\pm}=1+(2n_{\beta}+1)\ex{\pm 2 r}$.
The Fisher information can be computed from its definition in Eq.~\eqref{eq:fisher}
by taking the logarithmic derivative of the Gaussian \eqref{eq:hete-probability2},
\begin{equation}\label{eq:fisher-hete}
I=\int d\alpha'^2 \frac{\ex{- d^t O_{\phi}\Gamma_{S}^{-1} O_{\phi}d}}{\pi \sqrt{V_{+}V_{-}}} [ \partial_{\phi}( d^t O_{\phi}\Gamma_{S}^{-1} O_{\phi}d)]^2   .
\end{equation}
The derivative in the integrand is given by
$$ \frac{\partial{O_{\phi}\Gamma_{S}^{-1}O_{\phi}^t}}{\partial \phi} =\!-i O_{\phi}[\sigma_{y},\Gamma_{S}^{-1}] O_{\phi}^t=\left(\frac{1}{V_{-}}\!-\!\frac{1}{V_{+}}\right)
O_{\phi} \sigma_{x}  O_{\phi}^t,$$
where we have used $O_{\phi}=\exp (-i \phi \sigma_{y})$ and  $\sigma_{x}$ and $\sigma_{y}$ are the Pauli operators.
Making the change in variable $\alpha'\to \alpha'\ex{i \phi}$, which corresponds to $d\to O_{\phi}^t d$, equation \eqref{eq:fisher-hete} becomes
\begin{eqnarray}\label{eq:fisher-heteFIN}
I&=&\int d\alpha'^2 \frac{16 \alpha_{x}'^2 \alpha_{p}'^2}{\pi  \sqrt{V_{+}V_{-}}} \left(\frac{1}{V_{-}}\!-\!\frac{1}{V_{+}}\right)^2  \ex{-\frac{2 \alpha_{x}'^2}{ V_{-}}-\frac{2\alpha_{p}'^2}{V_{+}}}\nonumber\\
&=& \frac{(V_{-}-V_{+})^2}{ V_{-}V_{+}}=
\frac{4n_{r}(n_{r}+1)(2n_{\beta}+1)^2}{n_{r}(2n_{\beta}+1)+(1+n_{\beta})^2}
\end{eqnarray}

For pure states  the phase variance is given by
\be
\var_{\mathrm{sq}}^{\mathrm{het}}[\phi]=\frac{1}{4 N n_{r}} \hspace{1cm}  \mbox{ for }  \hspace{.2cm} n_{\beta}=0
\ee
which has a shot-noise $O(n^{-1})$ scaling, instead of the Heisenberg $O(n^{-2})$  scaling  that a squeezed vacuum state can provide:   \eqref{eq:varenen}.
However, from  \eqref{eq:fisher-heteFIN} we also notice that, as in the case of the optimal measurement, the Fisher information increases with temperature. In particular, in the limit of very high temperature  we find,
\be
\var_{\mathrm{sq}}^{\mathrm{het}}[\phi]=\frac{1}{16 N n_{r}(n_{r}+1)} \hspace{1cm}  \mbox{ for }  \hspace{.2cm} n_{\beta}\to\infty,
\ee
which coincides with the optimal bound  \eqref{eq:varenen}. So, heterodyning performs very poorly (sub-optimal scaling) for high-purity states; but quite surprisingly it attains the optimal bounds at high temperature.

Using \eqref{eq:par} and \eqref{eq:fisher-heteFIN} we can give the phase variance for an initial pure squeezed state of mean photon number $n_0$ that is sent through a lossy channel of transmittance $T$:
\be
\label{eq:var-heteT}
\Var_{\mathrm{sq}}^{\mathrm{het}}[\phi]=\frac{(1-R^2) n_0+1}{4 N T^2 n_0(n_0+1)},
\ee
where we recall that $R=1-T$.

In the limits of high and low initial squeezing we find,
\begin{eqnarray}
\Var_{\mathrm{sq}}^{\mathrm{het}}[\phi]&=&  \frac{2-T}{4 N T n_0} \hspace{1cm} \mbox{for }\; n_0\gg1,\\
\Var_{\mathrm{sq}}^{\mathrm{het}}[\phi]&=&\frac{1}{4 N T^2 n_0}
\hspace{.85cm} \mbox{for }\; n_0\ll 1.
\end{eqnarray}
Comparing with the optimal results in these regimes, \eqref{eq:ophs} and \eqref{eq:opls}, we see
that in the low squeezing regime heterodyning performs a factor 2 worse than the optimal strategy, but the difference becomes even more pronounced for high squeezing.

To calculate the  heterodyne variance for coherent states we proceed along the same lines. The outcome probabilities are now given by,
\begin{equation}
\label{eq:hete-probability}
 p(\alpha' |\phi)=1/\pi \; \tr [\ketbrad{\alpha'} \rho_{\beta,\alpha}(\phi)] =
\frac{\ex{-\frac{| \alpha'-\alpha\exp[i\phi]|^2}{n_{\beta}+1}}}{\pi \sqrt{n_{\beta}+1}} .
\end{equation}
From this the Fisher Information and the minimum variance can be readily computed:
\be
\Var_{\mathrm{coh}}^{\mathrm{het}}[\phi]=\frac{n_{\beta}+1}{2N|\alpha|^2}=\frac{n_{\beta}+1}{2N n_{\alpha}},
\ee
which again is sub-optimal for pure states (although only by a constant factor of 2)
and approaches the optimal bound \eqref{eq:varcoh} at high temperatures.

\subsection {Canonical phase-measurement}

In the one-copy case we saw that the phase measurement, defined after Eq.~\eqref{eq:boundfid}, was optimal. Although somewhat a theoretical exercise, it is interesting to study the accuracy of such measurement in the multicopy scenario. For pure squeezed states it is known that, although it is  sub-optimal,  the scaling of the variance with the mean photon number is the same as that for the optimal protocol: $\var[\phi]\sim 1/(N n^2)$. Here we would like to study its performance in the presence of losses. The computation for arbitrary squeezing is quite involved, but we can obtain analytical expressions in the two limiting cases of large and low initial squeezing.   From Eq.~\eqref{eq:encodest2} it is straightforward to get the expression of the probability,
\begin{align}\label{eq:prob-phase-multi}
p(\theta |\phi)&=p(\bar{\phi})=\frac{\sqrt{1-\lambda_{0}^2}}{2 \pi}\sum_{n,m=0}^{\infty}\left(\frac{\lambda_{0}}{2}\right)^{n+m}\ex{2i(n-m)\bar{\phi}}\nonumber\\
&\;\times \frac{\sqrt{(2n)!}\sqrt{2m!}}{n! m!}\sum_{k=0}^{\min (n,m)}\!\sqrt{p^{2n}_T(k)}\sqrt{p^{2m}_T(k)},
\end{align}
where $\bar{\phi}=\phi-\theta$ and $p^{2n}_T(k)$ are defined in \eqref{eq:binonomial}. For large squeezing the sum above is dominated by large values of $n$ and $m$. Then the binomial distribution $p^{2n}_T(k)$ is well approximated by a Gaussian distribution centered at $2nT$ and with variance $2nTR$, and the sum over    $k$ can be easily performed as an integral (for that we can extend the limits of integration to be $\pm\infty$). The  remaining factorials can be approximated using the Stirling formula as $\sqrt{(2n)!}/n!\simeq 2^n (n \pi)^{-1/4}$. We obtain
\begin{eqnarray}\label{eq:prob-phase-gauss}
p(\bar{\phi})&\simeq& \frac{\sqrt{1-\lambda_{0}^2}}{2 \pi}\sum_{n,m=0}^{\infty}\ex{2 i\bar{\phi}(n-m)}\lambda_{0}^{n+m}
\nonumber \\
&\times& \sqrt{\frac{2}{(m+n)\pi}} \ex{-\frac{(n-m)^2}{2(m+n)}\frac{R}{T}} .
\end{eqnarray}
At this point it is convenient to define the variables $s=m+n$ and $u=m-n$ and rewrite~\eqref{eq:prob-phase-gauss}
as
\begin{eqnarray}\label{eq:prob-phase-2}
p(\bar{\phi})&\simeq& \frac{\sqrt{1-\lambda_{0}^2}}{2 \pi}\sum_{s=0}^{\infty}\lambda_{0}^s \sqrt{\frac{2}{s \pi}}\int_{-\infty}^{\infty} du\;  \ex{-\frac{u^2 T}{2 s R}} \ex{2 i\bar{\phi} u}\nonumber \\
&=& \frac{\sqrt{1-\lambda_{0}^2}}{\pi}\sqrt{\frac{T}{R}}\sum_{s=0}^{\infty}\left( \ex{-2 \frac{T}{R}\bar{\phi}^2}\lambda_{0}\right)^s ,
\end{eqnarray}
where we have considered $u$ as a continuous variable and have extended the integration range from $(-s,s)$ to $(-\infty,\infty)$. This is legitimated as it amounts to neglecting terms that fall off exponentially in $u$.  The last sum can be trivially performed and yields
\begin{equation}\label{eq:prob-phase-3}
  p(\bar{\phi})\simeq\frac{\lambda_{0} \sqrt{1-\lambda_{0}^2} }{\pi}\frac{\sqrt{T/R}}{\ex{\frac{2 T}{R}\bar{\phi}^2}-\lambda_{0}}.
\end{equation}
Now we can  compute the Fisher information from Eq.~\eqref{eq:fisher} by noticing that for large squeezing ($\lambda_{0}\to 1$) the angular integral is dominated by small values of $\bar{\phi}$ and, therefore, $\exp[2(T/R){\bar{\phi}}^2]\simeq 1+ 2(T/R)\bar{\phi}^2$. Extending to $\pm\infty$  the range of the integral over $\theta$, and recalling that in this limit $\lambda_{0}\simeq 1-1/(2n_0)$, the Fisher information finally reads $I\simeq 2(T/R)n_0$. Thus,  the variance is
\begin{equation}\label{eq:var-phase-large}
 \var_{\mathrm{sq}}^{\mathrm{can}}[\phi] \simeq \frac{1-T}{2 NT n_0}.
\end{equation}
This value is again twice as large as the optimal variance  in~\eqref{eq:ophs}, and coincides with that of the heterodyne scheme.

We next focus on the phase measurements in the  opposite limit: $n_0\ll 1$. Here the computation is much less involved. One just needs to keep terms up to order $\lambda_{0}^2$  in Eq.~\eqref{eq:prob-phase-multi}, compute the Fisher information, and take its linearized expression to the same order $\lambda_{0}^2$. After performing the trivial angular integration, we obtain $I\simeq 4 T^2 \lambda_{0}^2\simeq 4 T^2 n_0$ (recall that in this limit $\lambda_{0}\simeq \sqrt{n_0}$). Therefore the variance is
\begin{equation}\label{eq:var-phase-low}
 \var_{\mathrm{sq}}^{\mathrm{can}}[\phi] \simeq \frac{1}{4 NT^2 n_0}.
\end{equation}
Note that this value is twice the optimal variance in the same limit \eqref{eq:opls}.

For coherent thermal states we proceed in a similar fashion. We first compute the phase-measurement outcome probabilities starting from the representation of displaced
thermal states \eqref{eq:rhocohtherm} for high-field amplitudes ($\alpha\gg 1$),
\begin{equation}\label{eq:prob-phase-cov}
p(\bar{\phi})=\frac{1}{\pi n_{\beta}}\int d\alpha'^2
\ex{-\frac{|\alpha'|^2}{n_{\beta}}} |S(\zeta \ex{i\varphi})|^2,
\end{equation}
where we have implicitly defined $\zeta \ex{i \varphi}\equiv (\alpha+\alpha')\ex{i  \bar{\phi}}$, and where
\begin{eqnarray}
S(\zeta \ex{i\varphi})&=&\frac{1}{\sqrt{2\pi}}\sum_{n}\ex{-\frac{\zeta^2}{2}}\frac{\zeta^n}{\sqrt{n!}}\ex{i \varphi n}\nonumber\\
&\approx& \frac{1}{\sqrt{2\pi}(2\pi\zeta^2)^\frac{1}{4}}\!\int_{-\infty}^\infty\!du  \ex{-\frac{(\zeta^2-u)^2}{4 \zeta^2}}\ex{i \varphi u} \nonumber\\
&=& \left(\frac{2 \zeta^2}{\pi}\right)^{\frac{1}{4}}\ex{-\zeta^2 \varphi^2}\ex{i\zeta^2 \varphi}.
\label{eq:S}
\end{eqnarray}
In the second equality we have used the Gaussian approximation to the Poissonian photon distribution of a coherent state and extended the lower limit of integration from zero to $-\infty$.
In the limit of large amplitudes $\zeta\approx \alpha \gg1$, and, hence, the distribution $|S(\zeta \ex{i\varphi})|^2$ is strongly peaked at $\varphi\approx 0$. Thus, the following approximation holds
\begin{eqnarray}
\zeta\varphi& \approx& \zeta \sin\varphi=\mathrm{Im}[(\alpha+\alpha')\ex{i  \bar{\phi}}]
\nonumber\\
&=& (\alpha+\alpha'_{x})\sin{\bar\phi}+ \alpha'_{y}\cos\bar\phi.
\end{eqnarray}
With this we can and easily perform the integral in  \eqref{eq:prob-phase-cov} since it becomes Gaussian. We obtain
\begin{equation}\label{eq:prob-covcoh}
p(\bar{\phi})=\sqrt{\frac{2\alpha^2}{\pi(2n_{\beta}+1)}}
\ex{-\frac{2\alpha^2\bar\phi^2}{2n_{\beta}+1}}.
\end{equation}

It is now is straightforward to perform the $\bar\phi$ integral in the definition of the Fisher information, Eq. \eqref{eq:fisher}, and obtain the variance for the phase-covariant measurement
\begin{equation}\label{eq:prob-covcoh2}
\Var_{\mathrm{coh}}^{\mathrm{can}}[\phi]=\frac{2n_{\beta}+1}{4N |\alpha|^2}=\frac{2n_{\beta}+1}{4N n_{\alpha}};
\end{equation}
which agrees with the optimal bound.

\subsection {Homodyne measurements}
For pure states the standard quantum limit is known to be attained by homodyne measurements with a simple two step adaptive protocol~\cite{monras_optimal_2006}.  These measurements are very relevant from the practical point of view since they can be readily implemented in optical experiments \cite{wiseman_adaptive_1995,armen_adaptive_2002}. It seems natural to expect that homodyne measurements are also  asymptotically optimal for thermal states, i.e., that they saturate the bound~\eqref{eq:varenen}. Let us anticipate that this is not the case.

In the limit of strong local oscillator, a homodyne measurement  is described by the set of projectors $\{\ket{x}_{\theta}\!\bra{x}\}$, where $\{\ket{x}_{\theta}\}$, are the eigenstates of the quadrature operator $\hat x_{\theta}=(a \ex{i\theta}+a^{\dag}\ex{-i\theta})/\sqrt{2}$. The probability of obtaining outcome $x$ upon measuring a shifted squeezed thermal state $\rho_{\beta,r}(\phi)$
is given by \cite{leonhardt_measuringquantum_1997},
\begin{eqnarray}\label{eq:homo-probability}
 p_{\theta}(x|\phi)&=&\bra{x}U(\phi-\theta)S(r)\rho_{\beta}S(r)\dg U(\phi-\theta)\dg\ket{x}
 \nonumber\\
 &=&\frac{1}{\sqrt{2 \pi \sigma^2(\theta-\phi)}}\exp[-\frac{x^2}{2 \sigma^2(\theta-\phi)}],
\end{eqnarray}
where
\begin{equation}\label{eq:sigma-homo}
 \sigma^2(\theta)=(2 n_\beta+1)(\ex{2r}\cos^2\theta+\ex{-2r}\sin^2\theta).
\end{equation}

The Fisher information can be easily computed from its definition in Eq.~\eqref{eq:fisher}. One obtains
\begin{equation}\label{eq:fisher-homo}
I_{\theta}(\phi)=\frac{1}{2}\left(\frac{\sin[2(\theta-\phi)](\ex{2r}-\ex{-2r})}{\ex{2r}\cos^2(\theta-\phi)+\ex{-2r}\sin^2(\theta-\phi)}\right)^2.
\end{equation}
At this point it is readily apparent that the temperature dependence has disappeared from the expression of the Fisher information.
The maximum of the $I_{\theta}(\phi)$  is achieved for an homodyning angle
\begin{equation}\label{eq:theta-homo-max}
 \theta^*=\phi\pm \arctan(\ex{2r}),
\end{equation}
and the maximum Fisher information reads
\begin{equation}\label{eq:fisher-homo-max}
 I^*=2 \sinh^2(2r)=8 n_r(n_r+1).
\end{equation}
Upon using the Cram\'{e}r-Rao bound we obtain that the maximum precision that can be achieved with homodyning measurements is
\begin{equation}\label{eq:var-homo}
\var_{\mathrm{sq}}^{\mathrm{hom}}[\phi]=\frac{1}{8N n_r(n_r+1)}.
\end{equation}
The above maximum of the Fisher information can only be attained if some previous information on the phase is available. It can be argued ~\cite{monras_optimal_2006} that one can use a vanishingly small fraction of copies to perform  a (perhaps) sub-optimal rough estimation of the phase and then use the remaining number of copies to nail down its precise value. The final estimation fidelity is dominated by this second stage, and therefore the above bound can be asymptotically attained. A similar analysis leading to the same results was recently carried out by Mitchell \ea  \cite{mitchell}.

Several comments are in order when comparing \eqref{eq:var-homo} with the optimal variance~\eqref{eq:varenen}. First, we observe that homodyning protocols are sub-optimal, they yield the optimal variance only for pure states. Second, we see that the variance \eqref{eq:var-homo} is independent of the temperature of the thermal state.
We can understand this result as follows.
For squeezed states, the effect of temperature is the same as rescaling the phase-space. At the same time, homodyne measurements can be understood in phase-space as projections of the distribution on one quadrature. Therefore, the effect of temperature in the measurement statistics can be accounted for by a simple rescaling, thus the final precision  is independent of the temperature.

The variance \eqref{eq:var-homo} for a squeezed state  that emerges from a lossy channel of transmittance $T$ can be calculated by using \eqref{eq:par},
\begin{equation}\label{eq:var-homo-loss}
 \var_{\mathrm{sq}}^{\mathrm{hom}}[\phi]=\frac{1+4 n_0 T(1-T)}{8NT^2n_0(n_0+1)},
\end{equation}
where we recall that $n_0=\sinh^2 r_0$ is the mean photon number of the initial squeezed vacuum that is sent through the channel. Of course, for $T=1$ one recovers the value of the variance for pure states~\cite{yurke_su2_1986, monras_optimal_2006}. For large squeezing, $n_0\gg 1$, one has
\begin{equation}
\label{eq:homo-large}
\var_{\mathrm{sq}}^{\mathrm{hom}}[\phi]=\frac{1-T}{2NT n_0},
\end{equation}
while for $n_0\ll 1$,
\begin{equation}
\label{eq:homo-low}
\var_{\mathrm{sq}}^{\mathrm{hom}}[\phi]=\frac{1}{8 NT^2 n_0}.
\end{equation}
So, in the presence of losses and for low squeezing, adaptive homodyning provides a variance a factor of 2 smaller than the variance of heterodyning, and attains the optimal bound \eqref{eq:ophs},  while for high squeezing homodyning provides half the precision of the optimal strategy but still outperforms heterodyning by a factor which depends on the amount of losses. 

For coherent thermal states, the same scaling argument we used for squeezed states does not hold because the displacement is not affected by an increase in temperature, while it is obviously affected by phase-space rescaling. It is however a straightforward exercise to check that the Fisher information for displaced thermal states is
\be
I_{\theta}(\phi)=\frac{4|\alpha|^2 \sin^2(\phi-\theta)}{2n_{\beta}+1}.
\ee
Therefore, the optimal variance  \eqref{eq:varcoh} for displaced thermal states can be attained with a one-step adaptive homodyne measurement by taking $\theta^*=\phi\pm\pi/2$, even for states with non-zero temperature,
\begin{equation}\label{eq:varcohhom}
\var_{\mathrm{coh}}^{\mathrm{hom}}[\phi]=\frac{1}{4N}\frac{2 n_{\beta}+1}{n_{\alpha}}.
\end{equation}
We notice here that the Fisher information, or the minimum  variance for that matter, are the same for a coherent state $\ket{\alpha}$ than for a displaced thermal state $D(\alpha')\rho_{\beta}D(\alpha')\dg$
with $|\alpha'|^2=|\alpha|^2 (2 n_{\beta}+1)$, which can be understood in terms of the scaling argument used above.

It still remains an open problem to find optimal and practical strategies that attain the bound for general squeezed thermal states.
We note that the performance of adaptive homodyne measurements is as good as, or better, than that of the single-shot canonical phase measurement for both displaced and squeezed thermal states.


\section{Frequency estimation}\label{sec:fe}

Sub shot-noise parameter estimation is one of the most important applications brought by the field of quantum information. Since the work of Huelga \ea \cite{huelga_improvement_1997} it is well-known that the use of exotic quantum states  provides a dramatic improvement on the estimation precision of several metrologically relevant parameters. It was immediately recognized that states that provide this enhancement can be specially fragile in the presence of decoherence or losses and that it is therefore mandatory  to asses whether the quantum advantage persists in noisy environments \cite{huelga_improvement_1997, andr_stability_2004}. Recently, there have been several in-depth studies on this issue in discrete systems, specially oriented to the so-called maximally path-entanglement state (NOON) state ($\ket{N,0}+\ket{0,N}$)\cite{shaji_qubit_2007,gilbert_use_2008,dorner_optimal_2008}.

Here we want to approach the problem for CV systems.
In particular we will use the previous minimum values for the phase estimation variance
to  study a model for frequency estimation in the presence of losses (which is the dominant source of errors in optical experiments).
We will study a scenario where the phase is imprinted at a given rate $\phi=\omega t $ on a Gaussian state, in a time interval $t$ during which the system suffers losses also at a constant rate $\eta$. Long times $t$ will increase the accumulated phase and hence improve the sensitivity of the measurement. However, the unavoidable losses will clearly limit the duration of the experiment. Our goal here is to find the optimal time $t^*$  to estimate the frequency $\omega$ with the highest precision.

As stated above, a BS can be used to model a lossy channel. In our scenario the input state, a pure squeezed or displaced state, is sent through a BS  with a time dependent transmittance $T$, $T=\ex{-\eta t}$. So, we can use the above results, \eqref{eq:varcoh} and \eqref{eq:varenen}, to optimize the time by taking into account that now the variance for a fixed number of copies $N$ is given by
\begin{equation}
\Var{[\omega]}=[Nt^2I]^{-1}.
\end{equation}

The case of coherent states can be solved analytically. When a coherent state $\ket{\alpha}$ is sent to a BS, the transmitted state is nothing but $\ket{T^{1/2}\alpha}$. So, we only have to optimize over time the following expression:
\begin{equation}
\var_{\mathrm{coh}}[\omega]=\frac{1}{4N|\alpha|^2\ex{-\eta t}t^2},
\end{equation}
where we have replaced $|\alpha|^2$ by $|\alpha|^2\ex{-\eta t}$ and taken the limit $\beta\to\infty$ in Eq. (\ref{eq:varcoh}). We find that the optimal time is $t^*=2/\eta$, which  gives the minimum variance
\be
\Var_{\mathrm{coh}}{[\omega]}=\frac{\ex{2} \eta^2}{16 N |\alpha|^2}.
\ee

\begin{figure}[htp]
{\includegraphics[width=3.3in]{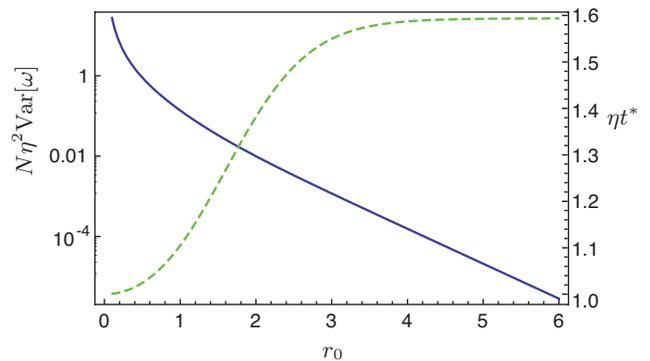}}
\caption{(Color online) Rescaled variance $N\eta^2 \var{[\omega]} $ (solid line) and optimal time $\eta t^*$ (dashed line) for the case of input squeezed vacuum state with squeezing parameter $r_{0}$.}\label{fig:los1}
\end{figure}

For squeezed input states we recall that the squeezing parameter and the temperature of the state at a given time $t$ depend on $T(t)=\ex{-\eta t}$ and on the initial squeezing~$r_{0}$ through Eq.~(\ref{eq:par}). Inserting this time dependence in \eqref{eq:optvarT} we can optimize $\Var_{\mathrm{sq}}{[\omega]}=\Var_{\mathrm{sq}}[\phi]/t^2$.
Figure \ref{fig:los1} shows the  optimal value of $\eta t^*$ for different values of $r_{0}$
together with the corresponding minimum variance.
The optimal time is now a function of the initial squeezing and ranges from $t^*=1/\eta$
for low squeezing to  $t^*=(2+W(-2 \ex{-2}))/\eta\approx 1.59/\eta$ for high squeezing, where $W(x)$ is Lambert's $W$-function \footnote{Lambert's W function satisfies $W(z)\ex{W(z)} = z$.}.  At this high squeezing regime the variance scales as $O(n_{0}^{-1})$. Hence,
the characteristic Heisenberg scaling of squeezed pure states turns into the standard quantum limit scaling in the presence of losses.
\begin{figure}[t]
{\includegraphics[width=3.3in]{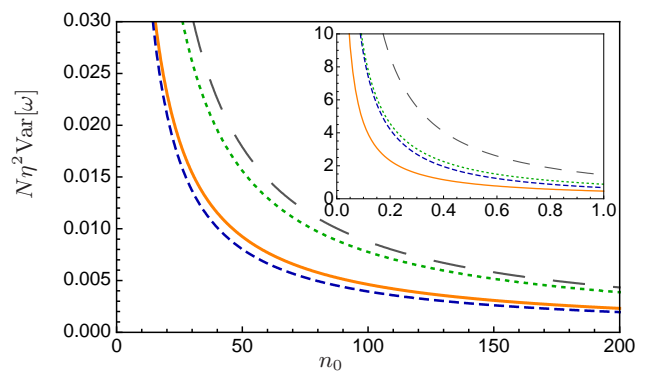}}
\caption{(Color
 online) Rescaled optimal variance $N\eta^2 \var{[\omega]}$ for  coherent (solid) and squeezed  (dashed) states  as a function of the mean photon number $n_{0}$ of the initial state.
The dotted line is the squeezed state variance for an adaptive homodyne measurement, while the long-dashed line is the variance for heterodyne measurements. Inset: As above but for a range  of small values of $n_{0}$.}\label{fig:cohsq}
\end{figure}

In Figure \ref{fig:cohsq} we compare the performance of an initial pure coherent state with that of a pure squeezed state of the same energy (mean photon number).
In the limit of low photon numbers (see inset) the squeezed state performs worse than
the corresponding coherent state of the same energy. At higher energies the squeezed state gives a slightly higher precision.

A similar behavior is found if one restricts to the adaptive homodyning strategy
mentioned above. In that case the variance in presence of losses for a given time $t$ can be computed from \eqref{eq:var-homo-loss}.
The optimal times  are shorter than those required for the optimal POVM, but they have the same asymptotic values  $\eta t^*\in[1,(2+W(-2 \ex{-2}))]$. Figure~\ref{fig:cohsq}  shows  the corresponding minimum variances for the homodyning strategy. We note that for small squeezing, homodyning is nearly optimal, while for large squeezing the differences become important (close to a factor of 2), though still give rise to the same type of scaling.

Finally, using \eqref{eq:var-heteT} we can obtain the results for the heterodyne strategy.
In this case the optimal times range from $\eta t^*=1$ for very low squeezing to $\eta t^*=2+W(-\ex{-2})\approx 1.84$. As shown in Figure~\ref{fig:cohsq}, heterodyning always performs worse than homodyning. The differences are especially large in the low squeezing regime, where homodyning is nearly optimal.

\section{Conclusions}
We have derived optimal bounds to the phase estimation fidelity for pure and mixed Gaussian states of light. This enables us to study how temperature affects the precision of the estimation. We have focused on displaced thermal and squeezed thermal states.  A priori, one would expect that the fidelity degrades with temperature, as the fluctuations also increase. This is precisely what we find for coherent thermal states. However, squeezed thermal states have the opposite behavior: a squeezed state with high temperature can provide twice the precision of a pure squeezed state with the same squeezing parameter.
The result seems at first sight paradoxical, but it can be understood
by noticing that an initial thermal state gives rise to a non-linear increase in the mean photon number,  which counterbalances the thermal noise.

In the many-copy scenario we have studied three different single-shot measurements:
heterodyning, canonical single-copy phase measurement, and adaptive homodyning.
Adaptive homodyne measurements attain the optimal bounds for coherent thermal states and squeezed vacuum states but are sub-optimal for thermal squeezed states since they provide a precision that is independent of temperature.
Heterodyning provides sub-optimal precision for pure states; however we find the surprising result that it does saturate the optimal bounds in the limit of very mixed states, a regime where adaptive homodyning ceases to be optimal.
Finally, what is taken to be the canonical phase measurement, which is optimal in the single-copy scenario but extremely hard to implement, turns out to perform worse than the practical heterodyne or adaptive homodyne strategies in some cases when many copies are available.

We have given a simple model to study the effect of losses in a situation where the phase imprinted on a gaussian state grows linearly with time, $\phi=\omega t$, while the system suffers losses also at constant rate $\eta$.  We give the optimal duration of the experiment which finds the compromise between losses and accumulated phase. Using the derived many-copy results and optimizing over this time interval,  we show that high-energy squeezed states give only a minor advantage over coherent states with the same energy,  and are in fact outperformed by them in the low energy regime.
While a similar analysis could be carried out for other Gaussian noisy channels (the complete Bures metric in the Gaussian state space, from which one obtains the optimal bound to the variance, can be found  in \cite{calsamiglia_quantum_2008}),
it remains an open problem to address optimal estimation for non-Gaussian noisy channels, such  as phase diffusion.

\vspace{-.30em}

\section{Acknowledgments}

\vspace{-.30em}

We are grateful to  Alex Monr\`{a}s  and Morgan Mitchell for discussions.  We
acknowledge financial support from the Spanish MICINN, through the
Ram\'on y Cajal program (JC), contracts FIS2005-01369,  and project QOIT
(CONSOLIDER2006-00019), from the Generalitat de Catalunya CIRIT 2005SGR-00994.


\end{document}